\documentstyle[twocolumn,aps,prl,epsfig]{revtex}

\begin{document}
\bibliographystyle{prsty}

\twocolumn[\hsize\textwidth\columnwidth\hsize\csname %
@twocolumnfalse\endcsname

\draft
\widetext

\title{Critical spin dynamics of the 2D quantum Heisenberg
antiferromagnets:\\ Sr$_2$CuO$_2$Cl$_2$ and Sr$_2$Cu$_3$O$_4$Cl$_2$}
\author{Y. J. Kim$^{1}$\cite{YJK}, R. J. Birgeneau$^{1,2}$, F. C. 
Chou$^1$, R. W. Erwin$^3$, and M. A. Kastner$^1$}
\address{$^1$Department of Physics and Center for Materials Science and
Engineering,\\
Massachusetts Institute of Technology, Cambridge, Massachusetts 02139}
\address{$^2$Department of Physics, University of Toronto, Toronto,
Ontario M5S 1A7, Canada}
\address{$^3$Center for Neutron Research, National Institute of
Standards and Technology, Gaithersburg, Maryland 20899}

\date{\today}
\maketitle

\begin{abstract}

We report a neutron scattering study of the long-wavelength dynamic spin
correlations in the model two-dimensional $S=1/2$ square lattice
Heisenberg antiferromagnets Sr$_2$CuO$_2$Cl$_2$ and
Sr$_2$Cu$_3$O$_4$Cl$_2$. The characteristic energy scale, $\omega_0
(T/J)$, is determined by measuring the quasielastic peak width in the
paramagnetic phase over a wide range of temperature ($0.2 \alt T/J \alt
0.7$). The obtained values for $\omega_0 (T/J)$ agree {\it quantitatively}
between the two compounds and also with values deduced from quantum Monte
Carlo simulations. The combined data show scaling behavior, $\omega \sim
\xi^{-z}$, over the entire temperature range with $z=1.0(1)$, in agreement
with dynamic scaling theory.

\end{abstract}

\phantom{.}
]

\narrowtext

Since the discovery of high temperature superconductivity in 1986 a great
number of studies have been devoted to understanding this fascinating yet
difficult problem. Although the superconducting mechanism itself in these
lamellar copper oxides is still elusive, tremendous progress in condensed
matter physics has resulted as a byproduct. One such example is our
understanding of the two-dimensional quantum Heisenberg antiferromagnet
(2DQHA) on a square lattice, which had been a long-standing problem in
theoretical physics even before the discovery of high temperature
superconductivity. Since the parent compounds of the copper oxide
superconductors, such as La$_2$CuO$_4$, are very good representations of the
$S=1/2$ square-lattice Heisenberg antiferromagnet, the study of the 2DQHA
has received renewed interest from both theorists and experimentalists
\cite{Kastner98}. The synergistic efforts of theoretical, numerical, and
experimental investigations have yielded a comprehensive picture of the
static magnetic properties of the 2DQHA over the past decade
\cite{Chakravarty89,Greven95a,Beard98}.

However, less progress has been made in studies of the dynamic properties
of the 2DQHA. In particular, there are only a small number of analytic
theoretical models and numerical simulations on this problem, while
systematic experimental investigation is lacking for the $S=1/2$ 2DQHA.
One of the most fundamental questions we can ask in the study of dynamic
critical behavior of the 2DQHA is whether or not dynamic scaling is
obeyed. The dynamic scaling hypothesis proposed by Ferrell {\it et al.}
\cite{Ferrell67} and developed for magnetic systems by Halperin and
Hohenberg \cite{Halperin69a}, is that the dynamic structure factor is
completely determined by the static properties with the scaling frequency
given by $\omega_0 \sim \xi^{-z}$. The theoretical value for the dynamic
critical exponent $z=d/2$ of the Heisenberg antiferromagnet has been known
for many years \cite{Hohenberg77}. Using this dynamic scaling hypothesis,
Chakravarty, Halperin, and Nelson (CHN) have argued that in the $S=1/2$
2DQHA the characteristic energy scale is given by $\omega_0 \propto
\xi^{-1} c_0 ( T / 2\pi \rho_s^0)^{1/2}$, where $c_0$ is the spin-wave
velocity and $\rho_s^0$ is the spin stiffness constant, both at zero
temperature \cite{Chakravarty89}. We set $\hbar=k_B=1$, so that
temperature $T$, and frequency $\omega$ have the units of energy.

The characteristic energy scale, $\omega_0$, for the 2DQHA is equivalent
to the relaxation rate of the order parameter, which can be determined by
measuring the damping of the quasielastic peak at the antiferromagnetic
wave vector $(\pi, \pi)$ via inelastic neutron scattering. One of the main
reasons for the scarcity of neutron experiments is the large energy scale
of the magnetism in the copper oxide materials. In magnetic systems, the
primary energy scale is set by the nearest neighbor exchange coupling $J$,
and in La$_2$CuO$_4$ or Sr$_2$CuO$_2$Cl$_2$ (2122) this is about $J\sim
130$ meV. Although thermal neutron scattering is one of the most powerful
tools in probing excitations in magnetic systems, it is not normally
suitable for studying the physics at such a high energy scale. As a
result, neutron scattering studies of the long-wavelength spin dynamics of
$S=1/2$ 2DQHA's such as La$_2$CuO$_4$ are very limited
\cite{Yamada89,Hayden90}.  The recent discovery of the system
Sr$_2$Cu$_3$O$_4$Cl$_2$ (2342), which contains a square lattice Cu$_{II}$
subsystem that has an order of magnitude smaller $J$, has opened the door
for a quantitative study of the spin dynamics of the $S=1/2$ 2DQHA
\cite{Kim99b}. In this Letter, we report systematic measurements of
$\omega_0$ as a function of $T/J$ in the model $S=1/2$ 2DQHA's, 2122 and
2342. We show that dynamic scaling is obeyed over the entire temperature
range probed in our experiment, $0.2 \alt T/J \alt 0.7$, which corresponds
to the correlation length range $100 \agt \xi/a \agt 2$.

\begin{figure}
\begin{center}
\epsfig{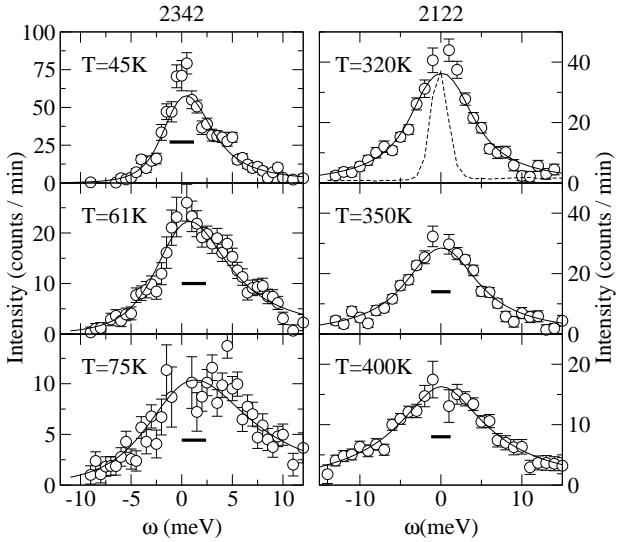}
\end{center}
\caption{
Representative $\omega$-scans at the $(\pi, \pi)$ position, which
corresponds to $(1/2 \; 1/2 \; 0)$ for both Sr$_2$Cu$_3$O$_4$Cl$_2$ and  
Sr$_2$CuO$_2$Cl$_2$, following the notation of Refs.
[3,11]. The background from incoherent scattering, measured far away
from the $(\pi, \pi)$ position, has been subtracted from the raw data. A
representative scan of the incoherent scattering, multiplied by 1/5, is
shown as a dashed line in the top right panel. A fixed final neutron
energy of 30.5 meV and collimations of 40'--48'--Sample--44'--40' have
been used. The solid lines are the results of least square fits to Eq.\
(\ref{eq:THC}) convoluted with the instrumental resolution function.}
\label{fig1}
\end{figure}

The neutron scattering experiments were carried out at the Center for
Neutron Research at the National Institute of Standards and Technology.
Large single crystal samples of both 2122 and 2342 were grown by the flux
method; typical sample dimensions were $20 \times 20 \times 5$ mm$^{3}$,
and the sample mosaicity was $\sim 0.2^\circ$ half width at half
maximum (HWHM). Various experimental configurations at the thermal
beamline BT9 were employed in the measurements. Progressively coarser
resolution was used to accommodate the wider energy width as the
temperature was increased and, correspondingly, $\xi$ decreased. Pyrolytic
graphite filters were used to reduce the effects of $\lambda/2$ neutrons.
The magnetic structure factor at $(\pi, \pi)$ as a function of energy
transfer, $\omega$, was measured at various temperatures.  The lower bound
of the temperature range was governed by the N\'eel ordering temperature
($T_N \approx 256$ K for 2122 and $T_N \approx 40$ K for 2342), below which 
the spin dynamics is no longer relaxational at $(\pi, \pi)$ 
\cite{Greven95a,Kim00b}. Typical
scans are shown in Fig.\ \ref{fig1} for both 2122 and 2342.  The energy
resolution is also plotted in each panel as a horizontal bar.

To determine $\omega_0$ these scans were fitted to the following
form for the dynamic structure factor 
convoluted with the instrumental resolution function:
\begin{eqnarray}
S({\bf q},\omega) &&= {\omega \over 1 - e^{-\omega/T}}{S(0) \over
1+(q\xi)^2} \left[{\Gamma_{\bf q} \over
(\omega-cq)^2 + \Gamma_{\bf q}^2} \right. \nonumber \\ 
&+& \left. {\Gamma_{\bf q} \over (\omega+cq)^2 +
\Gamma_{\bf q}^2} \right].
\label{eq:THC}
\end{eqnarray}
Here {\bf q} is defined as the deviation in wave vector from the 2D
antiferromagnetic position ($\pi,\pi$), and $c$ is the spin wave velocity.
We have used $\Gamma_{\bf q} \equiv \omega_0 [1 + \mu (q\xi)^2]^{1/2}$ as
proposed by Ty\v{c}, Halperin, and Chakravarty (THC) \cite{Tyc89}.  Our
fits turned out to be insensitive to the choice of the phenomenological
parameter $\mu$, and, accordingly $\mu$ was set equal to zero. In Eq.\
(\ref{eq:THC}), we define the characteristic energy scale as the HWHM of
the quasielastic peak, that is, $\omega_0 \equiv \Gamma_{{\bf q}=0}$. It
should be noted that Eq.\ (\ref{eq:THC}) is a simplified version of the
dynamic structure factor obtained by THC from their molecular dynamics
simulation of a classical lattice rotor model \cite{Tyc89}. Specifically,
we ignore the logarithmic corrections ($\log[1+(q\xi)^2]$) in the dynamic
structure factor of THC, since we only focus on the small ${\bf q}$
regime.

\begin{figure}
\begin{center}
\epsfig{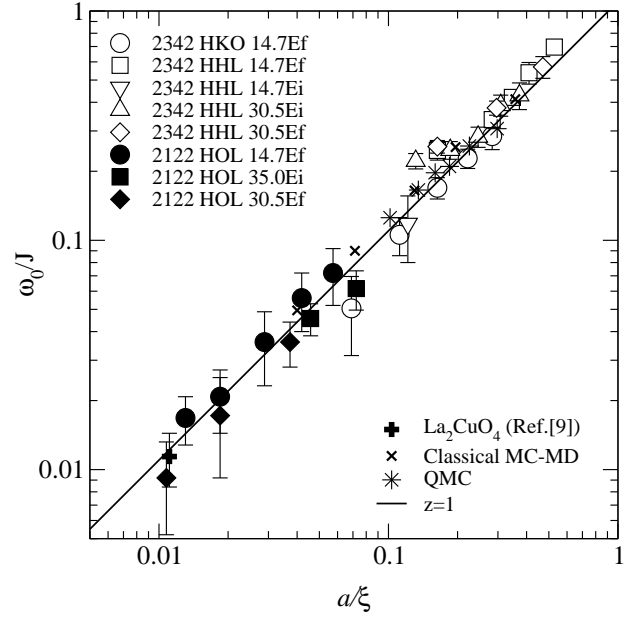}
\end{center}
\caption{Logarithmic plot of the reduced characteristic energy scale
$\omega_0/J$ versus the inverse of reduced spin correlation length
$a/\xi$. The open symbols are the data for Sr$_2$Cu$_3$O$_4$Cl$_2$ plotted
with $J=10.5$ meV, and the filled symbols are the data for
Sr$_2$CuO$_2$Cl$_2$ plotted with $J=125$ meV. Note that we use different
symbols for each experimental configuration. Here HKO denotes the
scattering plane in reciprocal space. We also plot the result of Ref. [9]
for La$_2$CuO$_4$ at $T=320$ K. The solid line is the dynamic scaling
prediction with $z=1$, that is, $\omega_0 \sim \xi^{-1}$. The results of
the quantum Monte Carlo simulations of the $S=1/2$ 2DQHA by Makivi{\'c}
and Jarrell {[13]} are plotted as $\ast$, while Wysin and Bishop''s
classical Monte Caro result {[14]} are plotted as $\times$ with the energy
scaled by $JS(S+1)$. }
\label{fig2}
\end{figure}

We plot the results for $\omega_0/J$ versus the inverse correlation length
$\xi^{-1}$ in Fig.\ \ref{fig2}. The values of the correlation length in
these systems have been determined previously with high accuracy
\cite{Greven95a,Kim00b}. Therefore, $\omega_0$ and the overall amplitude
$S(0)$ are the only adjustable parameters in the fits to Eq.\
(\ref{eq:THC}). In order to compare the two different systems, as well as
with the quantum Monte Carlo results, we scale $\omega_0$ by $J$, and
$\xi$ by the lattice constant $a$. Each data set obtained with a different
experimental setup is plotted in a different symbol. Filled and open
symbols are used for 2122 and 2342, respectively. In their study of the
spin dynamics of La$_2$CuO$_4$, Hayden and coworkers obtained $\omega_0$
at $T=320$ K \cite{Hayden90}, which is also plotted in Fig.\ \ref{fig2}.  
It is quite remarkable that the experimental results from all three
systems fall on a single straight line without any adjustable parameter.
This clearly shows the dynamic scaling behavior of $\omega_0$. If we fit
the data to the scaling form $\omega_0 \sim \xi^{-z}$, we obtain
$z=1.0(1)$, in excellent agreement with the theoretical value for the
2DQHA. The solid line is a fit of the experimental data to the scaling
form with fixed $z=1$: $\omega_0/J = 1.1(1) a\xi^{-1}$.

In order to compare our experimental results with the results of numerical
simulations, Makivi{\'c} and Jarrell''s quantum Monte Carlo results
\cite{Makivic92} for the $S=1/2$ 2DQHA are plotted in Fig.\ \ref{fig2}.
Wysin and Bishop's Monte Carlo molecular dynamics calculation results for
the classical 2D Heisenberg antiferromagnet are also plotted
\cite{Wysin90}. Not surprisingly, the quantum Monte Carlo results are in
an excellent agreement with our neutron scattering results. However, it is
interesting to note that the values from the classical Monte Carlo
calculation also agree with our neutron scattering results within
experimental error bars. Since Wysin and Bishop have calculated $\xi$ as
well as $\omega_0$, there is no adjustable parameter except the $J
\rightarrow JS(S+1)$ scaling. The good agreement between quantum and
classical dynamics presumably reflects the fact that the zero temperature
spin dynamics of the $S=1/2$ 2DQHA are well described by the classical
spin-wave picture with a uniform frequency renormalization of only 17\%.

Two important comments regarding the data analysis are in order. First,
due to the nonzero resolution width in both energy and momentum transfer,
scattered neutrons with ${\bf q} \neq {\bf 0}$ are invariably included in
our data shown in Fig. 1. However, from numerical simulations, we have
verified that the ${\bf q} \neq {\bf 0}$ dynamical contribution to the
total quasielastic scattering intensity for 2342 is insignificant and
yields a correction to the energy width that is much smaller than the
experimental error bars. On the other hand, 2122 has an order of magnitude
larger spin wave velocity (820 meV$\AA$) than that of 2342 (95 meV$\AA$),
and accordingly 2122 has a very steep dispersion relation. For 2122, the
nonzero ${\bf q}$ dynamical contribution thus gives an apparent peak width
in energy that is slightly larger than the intrinsic value. Although this
is already taken into account in our fitting processes, relatively large
error bars are generated for 2122 as a result. We have employed a number
of different experimental configurations to ensure that the extracted
$\omega_0$ is intrinsic in both 2122 and 2342. Second, we have defined the
characteristic energy scale as the HWHM of the dynamic structure factor at
${\bf q}={\bf 0}$. In Monte Carlo studies of Ref.\ \cite{Makivic92} and
Ref.\ \cite{Wysin90}, a different definition of $\omega_0$ was used, and
the values given in these studies are converted to fit our definition by
multiplying by a constant. We have obtained this constant multiplication
factor of 0.682 by fitting the raw data in Ref.\ \cite{Makivic92} to Eq.\
(\ref{eq:THC}).

The consistency among the experimental results and the numerical
simulation results in {\it absolute} units gives strong credence to the
combined data as a testing ground, against which various analytic theories
can be examined. Since the pioneering work by CHN, there have been a
number of theoretical studies to determine the characteristic energy scale
of the 2DQHA. They all seem to indicate $z=1$ dynamic scaling behavior of
$\omega_0$, but differ in the temperature dependence of $R_\omega$, which
is defined as a dimensionless ratio $R_\omega \equiv {\omega_0 \xi /
c_0}$. Specifically, $R_\omega \sim T^{1/2}$ is predicted by CHN, while
Grempel \cite{Grempel88} has carried out a conventional mode-mode 
coupling calculation and obtained the explicit prefactor of the $T^{1/2}$ 
dependence:
\begin{equation}
R_\omega \equiv {\omega_0  \xi \over c_0} = 1.3573 \left( {1 \over \pi}{1
\over Z_c^2 Z_\chi} {T \over J } \right)^{1/2} \approx 0.9 \sqrt{T \over
J}.
\label{eq:Grempel}
\end{equation}
This result is close to the expression obtained by THC in their numerical
simulation: $R_\omega=0.85 \sqrt{T/2 \pi \rho_s} \approx 0.8 \sqrt{T/J}$.  
However, Auerbach and Arovas, in their Schwinger boson mean field theory
\cite{Auerbach88}, have obtained a $T$-independent $R_\omega$.  Quantum
Monte Carlo calculations by Makivi{\'c} and Jarrell \cite{Makivic92} as
well as Nagao and Igarashi''s self-consistent theory \cite{Nagao98} also
show a $T$-independent $R_\omega$, although these results are limited to
a somewhat narrow temperature range.

\begin{figure}
\begin{center}
\epsfig{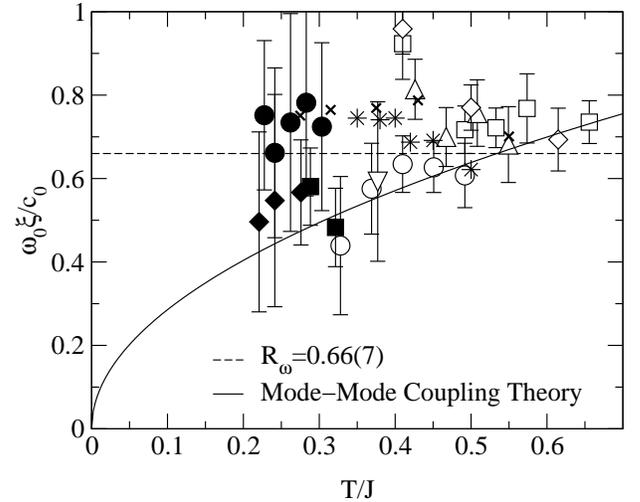}
\end{center}
\caption{The dimensionless ratio $R_\omega \equiv \omega_0 \xi /c_0$ as a
function of reduced temperature. The same symbols as in Fig.\ \ref{fig2}
are used in the plot. The solid line is the result from Grempel''s
mode-mode coupling theory, Eq.\ (\ref{eq:Grempel}). The dashed line is a
fit to a constant, $R_\omega = 0.66(7)$.}
\label{fig3}
\end{figure}

In Fig.\ \ref{fig3}, we plot the dimensionless ratio $R_\omega$ obtained
from our measurements as a function of reduced temperature. Note that
$c_0$ for the $S=1/2$ 2DQHA is well known from both Monte Carlo
\cite{Beard98} and series expansion studies \cite{Singh89a}:
$c_0=1.657Ja$. Therefore, Fig.\ \ref{fig3} also does not contain any
adjustable parameter. Results from Monte Carlo calculations are also
plotted in the figure, where the same symbols as in Fig.\ \ref{fig2} are
used. The analytic theory by Grempel, Eq.\ (\ref{eq:Grempel}), is plotted
as a solid line, while a fit to a presumed constant $R_\omega$ is plotted
as a dashed line.

We should emphasize that the dominant $T$ dependence of $\omega_0$ is
$\sim \xi^{-1}$, which has a strong exponential $T$ dependence, so that
Fig.\ \ref{fig3} only shows the weak $T$ dependence of the prefactor of
the leading $\xi^{-1}$ behavior. The salient feature to recognize in this
figure is that the coupled mode theory of Grempel is in a reasonable
agreement with our experimental results in {\it absolute} units.  
Considering that the mode-mode coupling approximation used by Grempel
ignores vertex renormalizations, the agreement in this intermediate
temperature range appears to be quite good. The large error bars on our
2122 data, however, prevent us from concluding that $R_\omega$ indeed has
the predicted $\sqrt{T}$ dependence. A temperature independent $R_\omega$
describes our data equally well, if not better. Another analytic theory
that predicts a $T$-independent $R_\omega$ is quantum critical theory
\cite{Sachdev00a}. However, quantitative comparison with our data is
difficult, since the temperature dependence of $\omega_0$ in the quantum
critical regime is not known theoretically.

It turns out that the task of experimentally determining the functional
form of $R_\omega$ is very difficult. Since $\sqrt{T}$ is a weak function
of $T$, one has to probe at very low temperatures to distinguish between
$\sqrt{T}$ and constant behavior. Although Sr$_2$CuO$_2$Cl$_2$ has the
smallest $T_N/J$ among model $S=1/2$ 2DQHAs known to this date, its large
nearest neighbor coupling makes a high-resolution neutron scattering study
very difficult. To elucidate further the temperature dependence of the
correction term $R_\omega$, a quantum Monte Carlo study of the spin
dynamics over a wide temperature range is highly desirable. It should be
noted that in the bicritical $S=5/2$ 2DQHA system Rb$_2$MnF$_4$,
$R_\omega$ appears to depend strongly on $T$; indeed in that case the
observed temperature dependence is much more rapid than $T^{1/2}$
\cite{Christianson01}.

In summary, a neutron scattering study of the long-wavelength dynamic spin
correlations in the model $S=1/2$ 2DQHA's Sr$_2$CuO$_2$Cl$_2$ and
Sr$_2$Cu$_3$O$_4$Cl$_2$ has been presented. We have measured the
characteristic energy scale as a function of temperature, and have shown
that dynamic scaling is valid over a wide temperature range ($0.2 \alt T/J
\alt 0.7$) with the dynamic critical exponent $z=1$. This temperature
range corresponds to the magnetic correlation length of $100 \agt \xi/a \agt
2$, for which dynamic scaling is expected to hold according to
theoretical predictions. Our attempt to determine the weak temperature
dependence of any possible corrections to this $z=1$ scaling is, however,
inconclusive, and remains to be addressed in future studies.

We would like to thank R. J. Christianson, M. Greven, Y. S.
Lee, R. L. Leheny, and S. Sachdev for helpful discussions. This work was
supported by the NSF award No. DMR0071256 and by the MRSEC Program of the
NSF under award No. DMR9808941 (at MIT), and by the NSF under agreement
No. DMR9423101 (at NIST).

\end{document}